\newif\ifiop
\ioptrue
\ifiop
  \documentstyle[epsf,rotate]{ioplppt}
	\def\putoffset{-1.7}
\else
  \documentstyle[aps,eqsecnum,epsf,rotate]{revtex}
	\def\putoffset{-1.5}
  \newdimen\mathindent
  \newcommand{\mathrm}{\rm}
  \newcommand{\e}{{\mathrm e}}
  
  \renewcommand{\i}{{\mathrm i}}
  
  \renewcommand{\d}{{\mathrm d}}
  \newcommand{\fl}{\hspace*{-\mathindent}}
  \newcommand{\tr}{\mathop{\mathrm tr}\nolimits}
  \newcommand{\opensqr}{\mbox{$\Box$}}
  \def\;{\protect\psemicolon}
  \def\psemicolon{\relax\ifmmode\mskip\thickmuskip\else\kern .3333em\fi}
  \makeatletter
    \newcommand{\opencirc}{\raisebox{2\p@}{\;\circle{5}}}
    \newcommand{\fullcirc}{\raisebox{-2\p@}{\Large$\bullet$}}
  \makeatother
  \newcommand{\fullsqr}{\mbox{\vrule height6pt width6pt}}
  \def\lo#1{\llap{${}#1{}$}}
\fi
\begin{document}
%----------------------------------------------------------------------
\ifiop
  \jl{1}
\fi
\title{Secular determinants of random unitary matrices}
\author{Fritz Haake\dag, Marek Ku\'s\dag\ddag,
  Hans--J{\"u}rgen Sommers\dag,
  Henning Schomerus\dag\ and Karol \.Zyczkowski\dag\S}
\address{\dag\ Fachbereich Physik, Universit{\"a}t GH Essen, 
  45117 Essen, Germany}
\address{\ddag\ Center for Theoretical Physics and College of Science,
  Polish Academy of Sciences, PL-02-668 Warszawa, Poland}
\address{ \S\ Instytut Fizyki im. Mariana Smoluchowskiego, Uniwersytet
  Jagiello\'nski, \\ ul. Reymonta 4,  PL-30-059 Krak\'ow, Poland}
\ifiop
  \relax 
\else
  \date{\today}
  \maketitle
  %\receipt{}
\fi
%----------------------------------------------------------------------
\begin{abstract}
  We consider the characteristic polynomials of random unitary
  matrices $U$ drawn from various circular ensembles. In particular, the
  statistics of the coefficients of these polynomials are studied. The
  variances of these ``secular coefficients'' are given explicitly for
  arbitrary dimension and continued analytically to arbitrary values of
  the level repulsion exponent $\beta$. The
  latter secular coefficients are related to the traces of powers of
  $U$ by Newton's well-known formulae. While the traces tend to have
  Gaussian distributions and to be statistically independent among
  one another in the limit as the matrix dimension grows large, the secular
  coefficients exhibit strong mutual correlations due to Newton's mixing of
  traces to coefficients. These results might become
  relevant for current efforts at combining semiclassics and
  random-matrix theory in quantum treatments of classically chaotic
  dynamics.
\end{abstract}
\ifiop
  \maketitle
\fi
%----------------------------------------------------------------------
\section{Introduction}
Circular ensembles of unitary matrices were first considered by
Dyson \cite{dys62} and described in detail by Mehta \cite{mehta}.
They are used to describe the quantum statistics of
periodically driven systems (see \cite{haake} and references therein)
and of scattering processes \cite{jp95}. It is the unitary Floquet
operator in the first case and the unitary $S$-matrix in the second
that the random unitary matrices in question attempt to mimic with
respect to certain more or less universal properties. Systems with
global chaos in their classical limit and with time reversal invariance
either present or strongly broken exhibit the greatest degree of
universality in their statistical properties and tend to fall in one of
the universality classes represented by the circular orthogonal,
symplectic, or unitary ensembles (COE, CSE, CUE). The so-called Poissonian
ensemble (CPE) of diagonal unitary matrices with independent unimodular
eigenvalues has also found applications for certain classically
integrable systems. There is a recent interest in the analysis of
intermediate
ensembles of unitary matrices which describe cross-over between
different universality classes \cite{ga70}, \cite{ps91}, \cite{mi95}.
Direct links between corresponding Gaussian and circular ensembles
have been established lately \cite{lewd91}, \cite{bro95}.

Unitarity constraints imposed on a random matrix of small size $N$
cause a significantly non-Gaussian character of the distribution of matrix
elements \cite{pm83}, \cite{mps85}, \cite{dsf91}. Moreover, various
statistics of unitary matrices from the different circular ensembles
depend strongly on the matrix size, and for small $N$ differ a great
deal from the asymptotic large $N$ properties. This is in contrast to
the Gaussian ensembles of Hermitian matrices \cite{mehta} which tend to
display lesser sensitivity to the matrix dimension.

For example, the normalized level spacing distribution $P(S)$ suffers a
cut-off at $S=N$, and is close to the Wigner-like distributions only
for $N$ upwards of roughly 10. Moreover, Baranger and Mello have
shown \cite{bm94} that the distribution of transmission intensities
$P(T)$ is non-Gaussian for $N=2$ and $4$, while coming close to
Gaussian for larger $N$.
%Several averages over the unitary group were computed by Mello \cite{m90}.

We were led to study the secular polynomials of random unitary matrices
in an attempt at constructing semiclassical quasienergy spectra for the
kicked top under conditions of classical chaos. Periodic-orbit theory
can be invoked only insofar as periodic orbits are available. In
practice, like for any system for which no simple symbolic dynamics is
known, one can hope to find all periodic orbits with periods no longer
than maybe $n_{\mbox{max}}\approx 10$, due to the infamous exponential
proliferation. These would allow to semiclassically evaluate traces of
powers of the Floquet operator, $\mbox{tr}F^n$, with $n$ up to
$n_{\mbox{max}}$. A Hilbert space of dimension
$N=2n_{\mbox{max}}$ is then accessible since Newton's formulae
\cite{Newton} (see Section \ref{s:coeff}) allow to express the first
$n_{\mbox{max}}$ coefficients of the secular polynomial in terms of
the first $n_{\mbox{max}}$ traces; the so-called
self-inversiveness of the secular polynomials of unitary matrices
\cite{lbb92} then yields the second half of the set of coefficients.
The practical applicability of periodic-orbit theory would thus seem
severely limited: Inasmuch as the dimension $N$ of the Hilbert space is
proportional to the effective size of quantum fluctuations (formally,
$N\propto 1/\hbar$) one runs out of periodic orbits just when the
semiclassical approximation begins to have a raison d'\^etre.

To ease the dilemma just described one might hope to increase the
size of the Hilbert space by throwing dice, according to random-matrix
theory, for traces with exponents $n_{\mbox{max}}<n<N-n_{\mbox{max}}$.
To prepare for such a ``marriage'' of semiclassical approximations with
random-matrix theory we here propose to study secular polynomials of
random unitary matrices from various ensembles. A previous first step in
this direction was taken in \cite{keh93}, where the means and the mean
squares of the coefficients mentioned were calculated for the CPE and the
CUE. A related study was presented by Bogomolny, Bohigas, and Leboeuf
\cite{lbb92}, \cite{lbb95} who gave the distribution of the roots of
random self-inverse polynomials.

Even though the bulk of the work to be presented is analytical we
performed extensive comparisons with numerical data on ensembles of
unitary matrices. To generate the data we constructed COE and CUE
matrices with the algorithm given in \cite{kzmk94} and CSE matrices as
described in \cite{kz95}.

\section{Coefficients of secular polynomials}
\label{s:coeff}
\subsection{Theory}

The secular polynomial of a unitary matrix $U$ of size $N$ is defined as
\begin{equation}
  \det(U-\lambda) = \sum^N_{n=0} (-\lambda)^n a_{N-n} =
  \prod_i \left(\e^{\i \varphi_i}-\lambda\right)
\label{F1}
\end{equation}
where the $\varphi_i$ are the eigenphases of $U$ and $a_0=1$. We are
interested in the statistics of the secular coefficients $a_n$ due to
various
ensembles of random matrices. We shall characterize these ensembles
by their joint densities of the $N$ eigenphases $\varphi_i$,
\begin{equation}
  d^\beta_N(\varphi_1,\varphi_2,\ldots \varphi_N) = C(\beta,N) \prod_{i<j}
  |\e^{\i\varphi_i}-\e^{\i\varphi_j}|^{\beta} ,
\label{d}
\end{equation}
where $C(\beta,N)$ is a normalization constant while $\beta$, the
so-called degree of level repulsion,
distinguishes the ensembles: $\beta=0$ for the
Poissonian ensemble for which the eigenphases are independently and
uniformly distributed over the interval $[0,2\pi)$; the circular
orthogonal, unitary, and symplectic ensembles are characterized by,
respectively, $\beta=1$, $\beta=2$, and $\beta=4$. Occasionally we shall
allow the parameter $\beta$ to range freely among the real numbers.
All of the ensembles in consideration are homogeneous inasmuch they do
not distinguish any particular value of any eigenphase. It follows
immediately that the ensemble means of all coefficients of the secular
polynomial vanish,
\begin{equation}
  \overline{a_n}=\int^{2\pi}_0\d\varphi_1\ldots\, \d\varphi_N a_N
  d_N^{\beta}(\varphi_1,\varphi_2,\ldots \varphi_N)=0.
\label{a_n}
\end{equation}

Next, we are interested in the ensemble averages $\overline{a_n a^*_m}$
which can be obtained from the generating function
\begin{eqnarray}
  P^{\beta}_N  & = & 
  \overline{\det(U-\lambda)(U-\mu)^\dagger} =
  {\overline{\prod_k(\e^{\i\varphi_k}-\lambda)(\e^{\i\varphi_k}-\mu)^*}}
\label{F2}\\
  & =&  \int^{2\pi}_0\d\varphi_1\ldots\, \d\varphi_N
  d^\beta_N(\varphi_1,\varphi_2,\ldots \varphi_N)
  \prod_k(\e^{\i\varphi_k}-\lambda)(\e^{\i\varphi_k}-\mu)^* \nonumber .
\end{eqnarray}
For $\lambda = \e^{\i\varphi},
\mu = \e^{\i\chi}$ it follows from the periodicity of the integrand that
the integral depends on the phases $\varphi,\chi$ only through the
variable $x=\e^{\i(\varphi -\chi)}$. Thus our generating function may be
written as
\begin{equation}
  P^{\beta}_N(x)=
  \overline{\prod_{i}(\e^{\i\varphi_i}-\e^{\i\varphi})(\e^{-\i\varphi_i}-
  \e^{-\i\chi})} =
  \overline{\prod_i f(\varphi_i,x)}=
  \sum^N_{n=0} x^n\overline{|a_n|^2}
\label{F3}
\end{equation}
where we have introduced the auxiliary function
\begin{equation}
   f(\varphi,x) =
   \left(\e^{\i\varphi}-x\right)\left(\e^{-\i\varphi}-1\right)\,.
\label{f}
\end{equation}
Hence all correlations
$\overline{a_n a^*_m}$ for $m\neq n$ vanish. Moreover, one easily shows
\begin{equation}
  \overline{|a_n|^2} = \overline{|a_{N-n}|^2}
\label{SI}
\end{equation}
which is in accord with the so-called
self-inversiveness, $a_{N-n}=a_n^*a_N$ \cite{Newton}; being a consequence
of but slightly weaker than unitarity self-inversiveness
entails each root of a
polynomial to either lie on the unit circle of the complex plane or to
be accompanied by its inverse as another root.

The variances $\overline{|a_n|^2}$ are most easily calculated in the
Poissonian case $\beta = 0$,
\begin{equation}
  P^0_N(x) = C(0,N) \left[\int^{2\pi}_0 \d\varphi\, f(\varphi,x)\right]^N
  = (1+x)^N .
\label{F4}
\end{equation}
Therefore $\overline{|a_n|^2} = {N\choose n}$ for $\beta = 0$ \cite{keh93}.

Next, we turn to the case $\beta = 2$ which was already
treated in reference \cite{keh93}. Observing that the
function $\prod_i f(\varphi_i,x)$ which we want to average is symmetric
in the $N$ phases $\varphi_i$ and that the density $d^2_N$ may be
written as a product of two Vandermonde determinants we have
\cite{mehta}, \cite{haake}
\begin{equation}
  P^2_N(x)=\int_0^{2\pi}\,\frac{\d^N\varphi}{(2\pi)^{N}}\,
  \det\left(\e^{\i(m-1)\varphi_m-\i(n-1)\varphi_m}\right)\,
  \prod_k f(\varphi_k,x)
\label{F5}
\end{equation}
with $m,n=1,2,\ldots N$ labelling the rows and columns of the determinant.
The integral over $\varphi_m$ can now be pulled into the $m$th row of
that determinant whereupon we immediately get
\begin{equation}
  \fl P^2_N(x) = \mbox{det}\left((1+x)\delta(m-n)-\delta(m-n+1)
   -x\delta(m-n-1)\right)
   =\sum^N_{n=0} x^N
\label{F6}
\end{equation}
where $\delta (m-n)$ denotes the Kronecker delta. The
foregoing generating function entails the variance
$\overline{|a_n|^2} = 1$ for $\beta = 2$.

A little more effort is required for the orthogonal and symplectic cases
$(\beta = 1,4)$. In the orthogonal case we again employ the symmetry of
the integrand of the $N$-fold integral in (\ref{F3}) to rewrite that
integral, which goes over the hypercube
$0\leq\varphi_1,\varphi_2,\ldots,\varphi_N\leq 2\pi$, as $N!$ times one
over the hypertriangle $2\pi>\varphi_1>\varphi_2>\ldots >\varphi_N>0$.
Within the hypertriangle the product of differences takes the form
\begin{equation}
  \prod_{k<\ell}|\e^{\i\varphi_k}-\e^{\i\varphi_{\ell}}| =
      \prod_{k<\ell} 2\sin\left(\frac{\varphi_k-\varphi_\ell}{2}\right) =
      \i^{\frac{N(N-1)}{2}}\det\left(\e^{\i m\varphi_1},\ldots\,,
      \e^{\i m\varphi_N}\right) ;
\label{F7}
\end{equation}
the second member in this chain of equated expressions is manifestly
positive whereupon the modulus operation can be dropped;
the determinant in the last member of (\ref{F7}) has its rows
labeled by the parameter $m$ which runs in integer steps between
$\frac{N-1}{2}$ and $\frac{N+1}{2}$; the label for the columns is the
one on the $N$ integration variables. We may thus write the generating
function as
\begin{equation}
\fl P^1_N(x) =N!\,C(1,N)\int\limits_{\varphi_1>\varphi_2\ldots>\varphi_N}
  \d\varphi_1\ldots \d\varphi_N
  \,\i^{\frac{N(N-1)}{2}}\det\left(\e^{\i m\varphi_1},\ldots,
  \e^{\i m\varphi_N}\right)\prod_{k} f(\varphi_k,x) .
\label{F8}
\end{equation}

For the symplectic case we find it expedient to extend the
$N$-fold integral in (\ref{F3}) to a $(2N)$-fold one,
\begin{eqnarray}
\fl P^4_N(x) = C(4,N)
  N!\,\int\limits_{\varphi_1>\varphi_2>\ldots>\varphi_{2N}}\d\varphi_1
  \ldots \d\varphi_{2N}\,
  \prod_{k<\ell}^{1\ldots 2N}\left(
  2\sin\left(\frac{\varphi_k-\varphi_\ell}{2}\right)\right)\nonumber\\
\label{F9}\\
  \times \prod_{k=1}^N
  \left[f(\varphi_{2k},x)\left(\frac{-\partial}{\partial\varphi_{2k-1}}
  \right)
  \delta\left(\varphi_{2k-1}-\varphi_{2k}-\epsilon\right)\right],\nonumber
\end{eqnarray}
where $\epsilon$ is to be sent towards zero from above. To see the
equivalence of the foregoing expression to $P^4_N$ as given by
(\ref{F3}) with $\beta=4$ we simply integrate by parts with respect to
the phases with odd labels. Only those terms survive for which the $N$
differentiations have turned precisely those $N$ sine functions into
cosines which are assigned vanishing arguments by the delta functions;
the remaining sine functions then come in quadruples like,
symbolically, $s_{13}s_{14}s_{23}s_{24}\rightarrow s_{24}^4$. We should
note in passing that the delta functions in the foregoing representation
of $P_N^4$ reflect Kramer's degeneracy and that in our definition of the
secular polynomial for the symplectic ensemble each of the 2-fold
degenerate eigenphases appears only once.

The integral representations (\ref{F8}) and (\ref{F9}) are convenient
starting points for an explicit evaluation of the generating functions
$P_N^1$ and
$P_N^4$. We propose to start with the slightly easier symplectic case.

In order to actually evaluate the $2n$-fold integral in (\ref{F9})
we once more employ the identity (\ref{F7}) as
extended to $2N$ ordered phases; we then integrate over every second
phase to get rid of the delta functions. Upon renaming the remaining
integration variables as
$\varphi_{2k}\rightarrow\varphi_k$ and letting the positive infinitesimal
$\epsilon$ go to zero we arrive at
\begin{equation}
\fl  P^4_N(x)
  = (-1)^N C(4,N)\,\int_0^{2\pi} \d^N\varphi
  \det\left(m\e^{\i m\varphi_1},
  \e^{\i m\varphi_1},m\e^{\i m\varphi_2},\e^{\i m\varphi_2},
  \ldots\right)
  \prod_{k=1}^N f(\varphi_k,x)
\label{F10}
\end{equation}
with $|m|\leq(2N-1)/2$. Note that we have returned to an integral over an
$N$ dimensional hypercube, exploiting the symmetry of the integrand.
The difficulty to be coped with now lies in the fact that each of the
$N$ integration variables appears in a pair of rows of the $2N\times
2N$ determinant. At
this point it is helpful to express that determinant
by a Gaussian integral over complex Grassmann variables as
\begin{equation}
  \det M =\int \left(\prod_{k} \d\eta_k^*\d\eta_k\right)
  \exp{\left(-\sum_{i,j}\eta_k^*M_{kl}\eta_l\right)}.
\label{Grass}
\end{equation}
Accounting for the matrix $M$ from (\ref{F10}) and doing the $2N$
integrals over the $\eta^*$ we arrive at
\begin{eqnarray}
  \int_0^{2\pi}\d^N\varphi\det M & = & \int \left(\prod_{k}\d\eta_k\right)
  \left[-\frac{1}{2}\sum_{m,m'}A_{m,m'}\eta_m \eta_{m'}\right]^N
  \nonumber\\
 & = & N! \int \left(\prod_{k} \d\eta_k\right)
  \exp{\left[-\frac{1}{2}\sum_{m,m'}A_{m,m'}\eta_m \eta_{m'}\right]}
  \label{Grass2}
\end{eqnarray}
with the antisymmetric $2N\times 2N$ matrix
\begin{eqnarray}
  \fl A_{mm'} = (m'-m) \int\limits_0^{2\pi}\d\varphi\,
    \e^{\i(m+m')\varphi}f(\varphi,x)\nonumber\\
  \lo= 2\pi(m'-m) \left((1+x)\delta(m+m')-\delta(m-m'+1)
     - x\delta(m+m'-1)\right);
\label{F12}
\end{eqnarray}
the labelling of the rows and columns of
$A$ is inherited from the ordered-phases form of the
Vandermonde determinant (\ref{F7}): Both $m$ and $m'$ run in integer
steps from $-(2N-1)/2$ to $(2N-1)/2$. The remaining Gaussian integral in
(\ref{Grass2}) is easily recognized as the Pfaffian $\sqrt{\det A}$ of the
antisymmetric matrix $A$, such that the generating function in search
takes the form
\begin{equation}
  P^4_N(x) = C(4,N) N!\, \sqrt{\det A}\,.
\label{F12a}
\end{equation}
The sign of the Pfaffian $\sqrt{\det A}$ must be chosen such that
$P^4_N(x)$ is positive for positive values of $x$, according to the
definition of the generating function.

In a first attempt at evaluating the Pfaffian one may rejoice in the
ease in finding it for small values of $N$ which latter suggest a
surmise for the general variance,
\begin{equation}
  \overline{|a_n|^2} = {N\choose n}\,\frac{1\cdot 3\ldots(2n-1)}
  {(2N-1)(2N-3)\ldots(2N-2n+1)} ,\qquad \mbox{for}\, \beta=4.
\label{F13}
\end{equation}
This conjecture will be proven in the appendix.

We finally turn to the orthogonal case, taking up the integral
representation (\ref{F8}) for $P^1_N(x)$. For the sake of
simplicity let us assume an even dimension $N$.
We start with integrating over the angles $\varphi_i$
with even indices $i$, pulling the $(2k)$th such integral into the
$(2k)$th
row of the determinant; while that integral at first appears as going
over the interval $\varphi_{2k+1}<\varphi_{2k}<\varphi_{2k-1}$ we can
hurry to replace the lower limit with 0, simply by adding the $N$th
column to the $(N-2)$th, the resulting $(N-2)$th to the $(N-4)$th
and so forth and thus obtain
\begin{eqnarray}
\fl P^1_N(x)
 =  C(1,N)N!\,\int\limits_{\varphi_1>\varphi_3>\ldots>\varphi_{N-1}}
            \d\varphi_1 \d\varphi_3\ldots
            \d\varphi_{N-1}\,\i^{\frac{N(N-1)}{2}}
            \label{F14} \\
\fl \qquad\times\det\left(\e^{\i m\varphi_1}f(\varphi_1,x),
\int\limits^{\varphi_1}_0 \d\varphi\, \e^{\i m\varphi}f(\varphi,x),\,
\e^{\i m\varphi_3}f(\varphi_3,x),\int\limits^{\varphi_3}_0 \d\varphi\,
\e^{\i m\varphi}f(\varphi,x),\ldots\right) . \nonumber
\end{eqnarray}
Now the integrand is symmetric in the remaining $N$ integration
variables whereupon we may extend the integration range to the
$N$ dimensional hypercube of edge length $2\pi$ and make up by dropping
the factor $N!$, as we had previously done in
(\ref{F10}); the analogy with (\ref{F10}) in fact goes much further:
Once more, every integration variable appears in two rows of the
determinant. Going through precisely the same reasoning as before we
again incur a Pfaffian form,
\begin{equation}
  P^1_N(x) = C(1,N)N!\,\sqrt{\det B}
\label{F15}
\end{equation}
with the slightly more unpleasant antisymmetric matrix
\begin{equation}
  B_{mm'}= -\i
  \int\limits^{2\pi}_{0}\d\varphi\,\d\varphi'f(\varphi,x)f(\varphi',x)
     \e^{\i(m\varphi+m'\varphi')}\rm{sign}(\varphi-\varphi')\,;
\label{F16}
\end{equation}
here, the sign function ensures the antisymmetry of the matrix $B$ with
$|m|\leq (N-1)/2$. Again, the Pfaffian suggests, by its easily
evaluated form for small dimensions $N$, a guess for the
variances,
\begin{equation}
\overline{|a_n|^2}=1+\frac{n(N-n)}{N+1} \qquad \mbox{for}\,\beta=1.
\label{F17}
\end{equation}
We refer to the appendix for the proof of that surmise for $N$
even or odd.

Upon inspecting the variances found above for $\beta=0,2$ and
conjectured for $\beta=1,4$ we were led to extrapolate to arbitrary
non-negative $\beta$ as
\begin{eqnarray}
 \overline{|a_n|^2} & =& 
  {N\choose n}\,\frac
  {1\cdot(1+\frac{\beta}{2})(1+\beta)
    \ldots\left(1+\frac{(n-1)\beta}{2}\right)}
  {\left(1+(N-1)\frac{\beta}{2}\right)\left(1+(N-2)\frac{\beta}{2}\right)
    \ldots\,\left(1+(N-n)\frac{\beta}{2}\right)}
  \nonumber\\
  &=&  {N\choose n}
  \frac{\Gamma(n+\frac{2}{\beta})\Gamma(N-n+\frac{2}{\beta})}
  {\Gamma(\frac{2}{\beta})\Gamma(N+\frac{2}{\beta})}.
\label{F18}
\end{eqnarray}
This expression has poles at $\beta=\frac{-2}{N-1},
\frac{-2}{N-2},\ldots\,\frac{-2}{N-n}$ and zeros at $\beta =
\frac{-2}{1}, \frac{-2}{2}, \frac{-2}{3},\ldots\,\frac{-2}{n-1}$ for
$n\leq\frac{N}{2}$. It is thus analytic and positive for all positive
$\beta$,
and it goes to zero for $\beta\rightarrow\infty$. This looks like strong
evidence for the general validity claimed before. We shall actually turn
the conjecture into a theorem in the appendix.
The proof will be based on the
fact that (\ref{F18}) is equivalent to the differential equation
\begin{equation}
\fl  \frac{\partial}{\partial
  x}\left(1+\frac{\beta}{2}(N-x\,\frac{\partial}{\partial x})\right)
  P^{\beta}_N(x) = \left(N-x\,\frac{\partial}{\partial x}\right)
  \left(1+\frac{\beta}{2}\,x\,\frac{\partial}{\partial x}\right)
  P^{\beta}_N(x)
\label{F19}
\end{equation}
for the generating function $P^{\beta}_N(x)$.

\subsection{Numerical results}

We constructed random unitary matrices $U$ of different sizes according
to the algorithm developed in \cite{kzmk94} for the CUE and the COE and
later generalized for the CSE \cite{kz95}. For each such matrix we
calculated a complete set of $N$ secular coefficients $a_n$ by first
computing the traces of arbitrary powers, $t_n=\tr (U^n)$, via
either matrix multiplication or diagonalization; Newton's formulae
\cite{Newton} (see next section) then led to the $a_n$.

Precise estimates of the variance of any random variable require a much
larger sample than estimates of the mean. We therefore present numerical
results obtained for large samples of relatively small random matrices
($N\sim 20$), although some computations done for $N\sim
200$ provide similar results.

\begin{figure}
\unitlength1cm
\begin{picture}(8,5)
\epsfxsize8cm
\put(2,\putoffset){\rotate[r]{\epsfbox{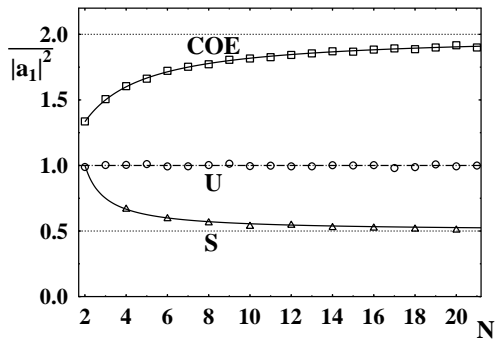}}}
\end{picture}
\unitlength1bp
\caption{Variance of the distribution of the first coefficient
  $\overline{|a_1|^2}$ as a function of the matrix size $N$ for
  circular ensembles: COE (\protect\opensqr), CUE (\protect\opencirc)
  and CSE $(\triangle)$. Solid lines
  represent analytical results and dotted lines the asymptotic behaviour.}
\label{f1}
\end{figure}
\begin{figure}
\unitlength1cm
\begin{picture}(8,5)
\epsfxsize8cm
\put(2,\putoffset){\rotate[r]{\epsfbox{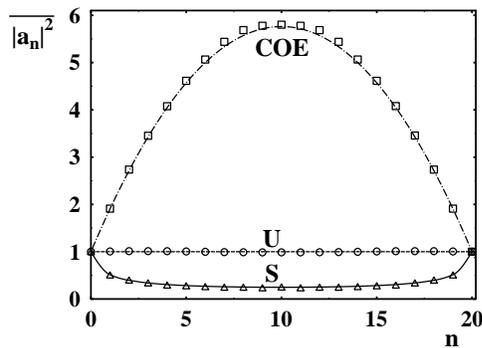}}}
\end{picture}
\unitlength1bp
\caption{Variance of the distribution of the $n$th coefficient
  $\overline{ |a_n|^2}$ obtained from $40000$ matrices of size $N=20$
typical of COE (\opensqr), CUE (\protect\opencirc) and CSE $(\triangle)$.}
\label{f2}
\end{figure}

Our above formulae for the variance $\overline{|a_n|^2}$ of the secular
coefficients
involve the index $n$ of the coefficient and the matrix size $N$.
Figure \ref{f1} represents the variance of the first coefficient as a
function of $N$, while Figure \ref{f2} shows the dependence on $n$ for a
fixed matrix size. Due to the property (\ref{SI}) of self-inversiveness
the latter curve is symmetric about $n=N/2$.

The data of all three ensembles coincide (up to a statistical
error) with the theoretical predictions. Note that the width of the
distribution of coefficients decreases with increasing
degree of repulsion and is smallest for the symplectic ensemble.
Additional numerical investigations confirmed the expectation that the
phases of the $a_n$ are distributed uniformly in the range
$[0,2\pi)$ for any canonical ensemble.
\begin{figure}
\unitlength1cm
\begin{picture}(8,5)
\epsfxsize8cm
\put(-1,\putoffset){\rotate[r]{\epsfbox{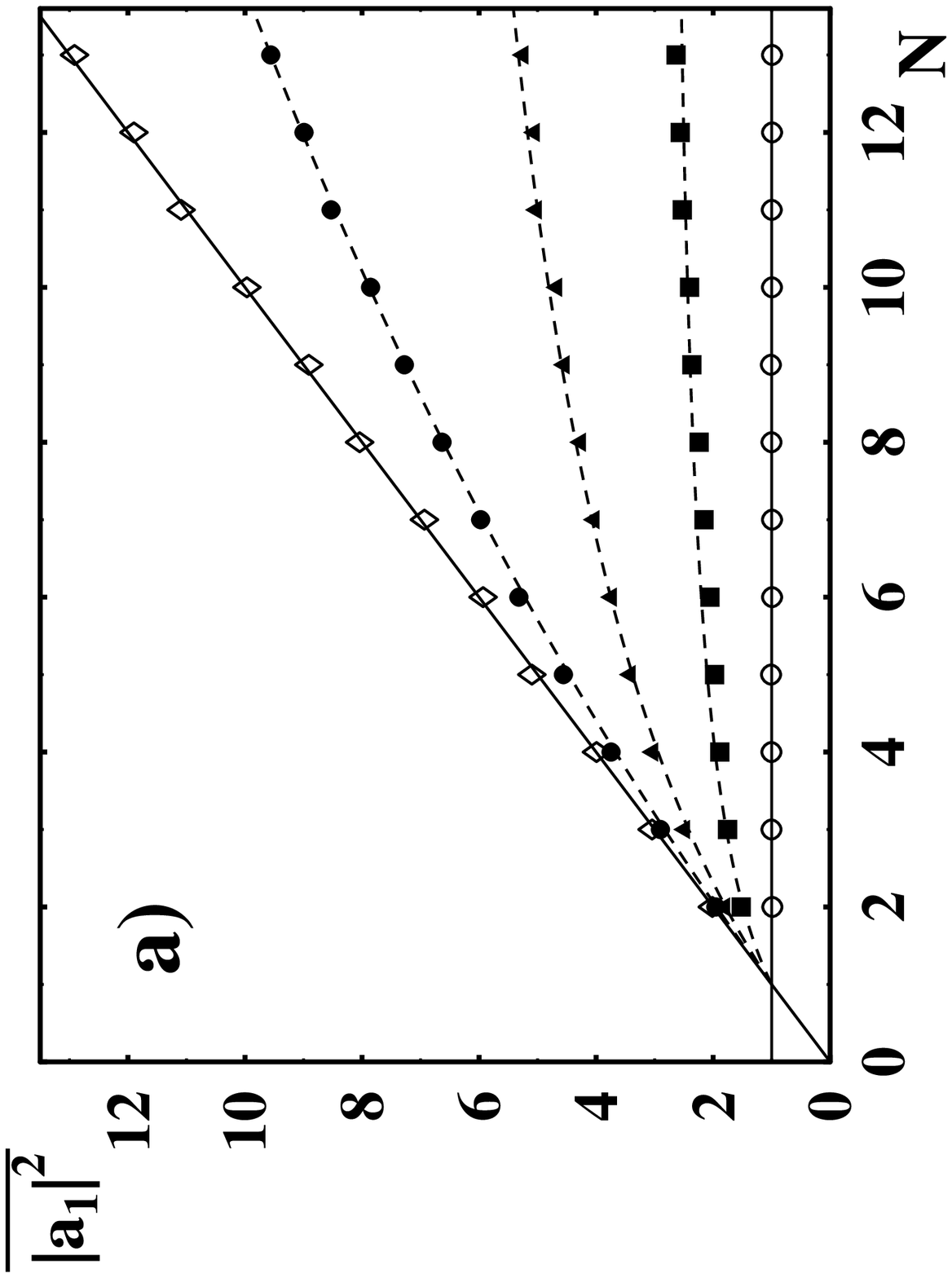}}}
\put(6.5,\putoffset){\rotate[r]{\epsfbox{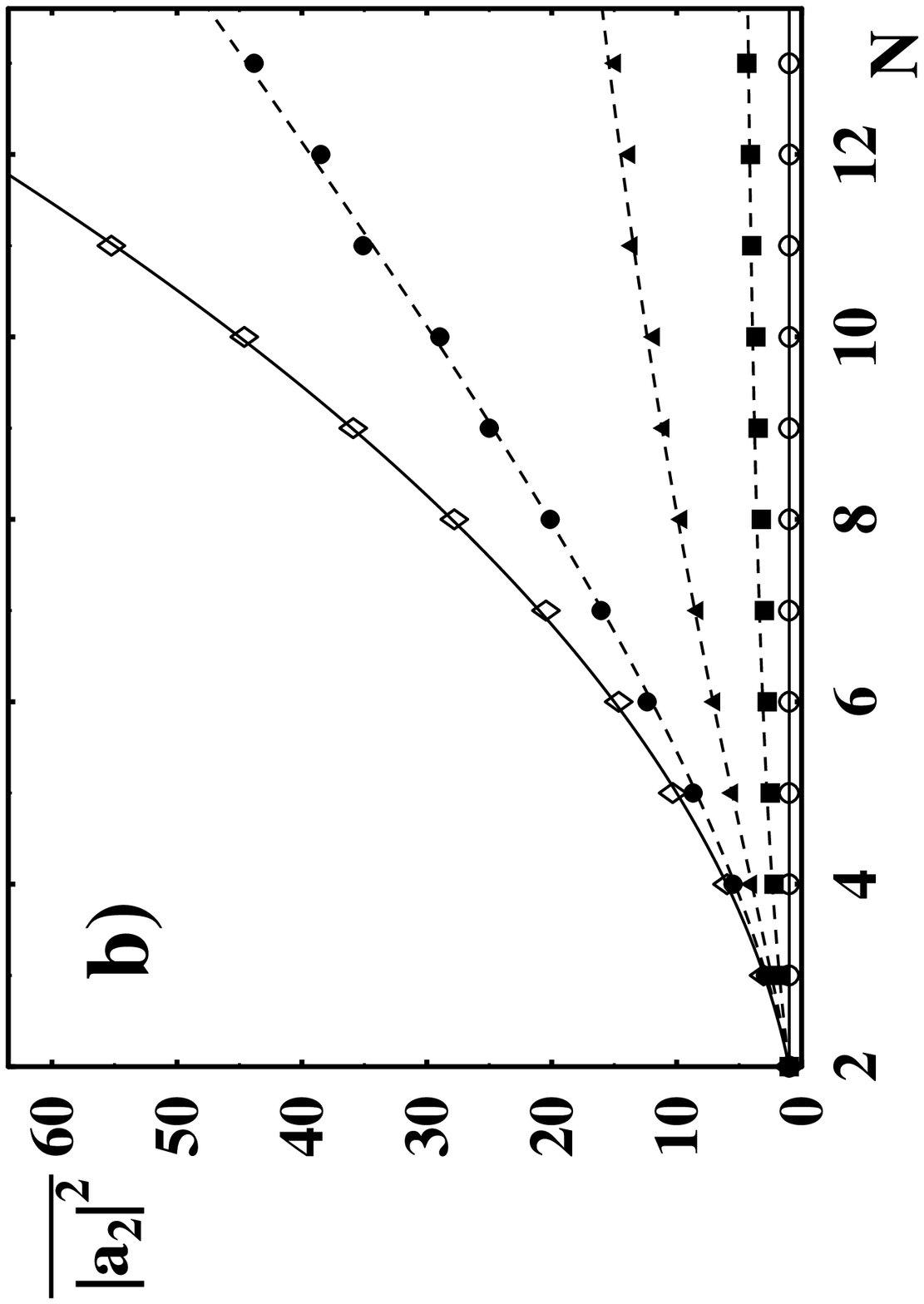}}}
\end{picture}
\unitlength1bp
\caption{Variance of coefficients for cross-over CPE - CUE:
  $\overline{ |a_1|^2}$ (a) and $\overline{ |a_2|^2}$ (b) as
  functions of the matrix size $N$. Control parameter
  $\delta$ is equal to $0.0$ (Poisson, $\Diamond $),
  $0.2$ (\protect\fullcirc), $0.4$, $0.7$ (\fullsqr), and
  $1.0$  (CUE, \protect\opencirc), while the level-repulsion parameter
  $\beta$ fitted simultaneously for {\sl both}
  coefficients equals $0.0, 0.06, 0.24, 0.69$, and $2.0$, respectively.}
\label{f3}
\end{figure}

Even though the formula (\ref{F18}) for the variance is primarily meant
to cover the four canonical ensembles $\beta=0$ (CPE), $1$ (COE), $2$
(CUE), and $4$ (CSE), we could not resist the temptation to test its
usefulness for intermediate cases. To this end we constructed an
ensemble of unitary matrices interpolating between the Poisson and
unitary ensembles according the method presented in \cite{kzmk95}. This
intermediate ensemble depends on one control parameter $\delta$, varying
from $0$ (CPE) to $1$ (CUE). Figure \ref{f3} shows the dependence of the
variance of the first two coefficients on the matrix size $N$ for
$\delta$ equal to $0.0, 0.2, 0.4, 0.7$, and $1.0$. For each case the
value of the parameter $\beta$ chosen to fit the $N$ dependence of
$\overline{|a_1|^2}$, inserted into (\ref{F18}) provides a fair
approximation for $\overline{|a_2|^2}$, and similarly for subsequent
coefficients. This astonishing fact reveals a certain validity of the
general formula (\ref{F18}) with non-integer values of $\beta$ for
ensembles in between the usual universality classes.

\section{Traces of powers of matrices from circular ensembles}
\label{s:traces}

The characteristic polynomial of a matrix $U$ is related to the traces
of its powers
\begin{equation}
  t_n = \tr (U^n) = \sum_{i=1}^N \e^{\i n\varphi_i}
\label{F20}
\end{equation}
by
\begin{equation}
  \det(U-\lambda)
  = (-\lambda)^N\exp\left(\tr \ln(1-\frac{1}{\lambda}U)\right)\,.
\label{F19a}
\end{equation}
Expanding both sides in powers of $\lambda$ one finds the
explicit relations
between the $a_n$ and the $t_n$ which were already established by Newton
\cite{Newton}. A compact representation is
\begin{equation}
  a_n=\frac{1}{n!}\left|\begin{array}{llllcl}
  t_1 & 1   & 0   & 0 & \cdots & 0\\
  t_2 & t_1 & 2   & 0 & \cdots & 0\\
  t_3 & t_2 & t_1 & 3 & \cdots & 0\\
  \multicolumn{6}{c}{\dotfill}    \\
  t_n & t_{n-1} & t_{n-2} & t_{n-3} & \cdots & t_1
  \end{array}\right| .
  \label{Newton}
\end{equation}
We infer from (\ref{F19a}) that there are only $N$ independent traces
$t_1,\ldots,t_N$. Moreover, for unitary matrices for which the
eigenphases
$\varphi_i$ are real the number of independent
complex parameters is again reduced by a factor 1/2: The first $N/2$
traces suffice to determine all $N$ coefficients $a_n$ (cf (\ref{SI})).

Clearly, the $t_n$ all vanish in the mean for all of the circular
ensembles considered here, due to the uniformity of the distribution of
the $\varphi_i$. For a more complete characterization of the statistics
of the traces we propose to calculate their marginal probability densities
\begin{equation}
P_{N,n}^{\beta}(t) = \overline{\delta^2\left(t-\sum_{i=1}^N
\e^{\i n\varphi_i}\right)}
\label{F21}
\end{equation}
where $\delta^2(t)$ is a two dimensional delta function in the complex
$t$ plane; the ensemble average is to be done with the weight (\ref{d}).
It turns out convenient to first calculate the Fourier transform
\begin{equation}
  \widehat{P}_{N,n}^{\beta}(k) =
  \overline{\exp\left(-\frac{\i}{2}\sum_i(k\e^{-\i n\varphi_i}+
  k^*\e^{+\i n\varphi_i})\right)}\,.
\label{F22}
\end{equation}
Due to the periodicity of all functions of the phases $\varphi_i$
involved the characteristic function $\widehat{P}_{N,n}^{\beta}(k)$
depends on $k$ only through the modulus $|k|$ while the density
$P_{N,n}^{\beta}(t)$ is only a function of $|t|$. Henceforth we assume
$k = k^* = |k|$ and write
\begin{equation}
  \widehat{P}_{N,n}^{\beta}(k) =
  \overline{\exp\left(-\i k\sum_i\cos(n\varphi_i)\right)}\,.
\label{F23}
\end{equation}

We immediately obtain for the Poissonian ensemble
\begin{equation}
  \widehat{P}_{N,n}^0(k) = J_0(k)^N ,
\label{F24}
\end{equation}
$J_0(k)$ being a Bessel function. It follows that the densities
$P^0_{N,n}$ are the same for all values of the exponent $n$, a rather
intuitive result given the statistical independence of the phases
$\varphi_i$ in the Poissonian case. Equally expected for such a
Poissonian random walk is the independence of the mean squared
``displacement'' of $n$, $\overline{|t_n|^2}=N$.

For the unitary ensemble, $\beta = 2$, we exploit the symmetry of the
exponential to be averaged in the phases $\varphi_i$ and employ the
analogue of (\ref{F5}) and find the characteristic function to take the
form of a Toeplitz determinant,
\begin{equation}
  \widehat{P}_{N,n}^2(k) =
  \det\left(\int_0^{2\pi}\frac{\d\varphi}{2\pi}\,\e^{\i\varphi(\ell-m)}\,
  \e^{-\i k\cos(n\varphi)}\right),\qquad \ell,m=1,\ldots N.
\label{F25}
\end{equation}
The remaining $\varphi$ integral again
yields Bessel functions and can be written as
\begin{equation}
  \widehat{P}_{N,n}^2(k) =
  \det\left(\sum_{s=-\infty}^{+\infty}J_{|s|}(k)(-\i)^{|s|}
  \delta(\ell-m+ns)\right).
\label{F26}
\end{equation}
Clearly, the number of non-zero elements of the $N\times N$ determinant
here incurred decreases as the order $n$ of the trace $t_n$ in
consideration grows. In particular, for $n\geq N$ only the diagonal
elements are non-zero such that
\begin{equation}
  \widehat{P}_{N,n}^2(k) = J_0(k)^N\qquad \mbox{for}\quad n\geq N.
\label{F27}
\end{equation}
At the other extreme, $n=1$, we meet with the full Toeplitz determinant
\begin{equation}
  \widehat{P}_{N,1}^2(k) = T_N = \left|
  \begin{array}{cccc}
    J_0    & -\i J_1 & -J_2 & \cdots\\
    -\i J_1  & J_0   & -\i J_1 & \cdots\\
    -J_2   & -\i J_1 & J_0  & \cdots\\
    \vdots &       &      &
  \end{array}
  \right|
\label{F28}
\end{equation}

Intermediate values of $n$ lead to the subdeterminants $T_m$ obtained by
cancelling the last $N-m$ rows and columns of $T_N$,
\begin{eqnarray}
  \widehat{P}_{N,n}^2(k) = T_2^{N-n} \cdot T_1^{2n-N}
  \qquad &\mbox{for} \quad N\geq n\geq\frac{N}{2}, \nonumber \\
  \widehat{P}_{N,n}^2(k) = T_3^{N-2n}\cdot T_2^{3n-N} \qquad
  & \mbox{for} \quad\frac{N}{2}\geq n\geq \frac{N}{3}
\label{F30}
\end{eqnarray}
and so forth. This can be seen as follows. Starting with
$N\geq n\geq\frac{N}{2}$ one checks that the determinant in (\ref{F26})
has non-vanishing elements residing only in the diagonal,
$J_0(k)\delta(\ell-m)$, and in two subdiagonals,
$-\i J_1(k)\delta(\ell-m+n)$.
One moves the $(1+n)$th row to become the second, then the $(1+n)$th
column to become the second and thus isolates a $2\times 2$ block
$T_2=J_0(k)^2+J_1(k)$ in the upper left corner. One repeats this process
by moving the $(2+n)$th rows and columns to become the fourth and so forth
until one arrives at a block
diagonal determinant in which the $2\times 2$ block
$T_2$ and the $1\times 1$ block $T_1=J_0(k)$ appear
$N-n$ times and $2n-N$ times, respectively. The procedure for
$\frac{N}{2}\geq n\geq \frac{N}{3}$ is analogous: One moves the
$(1+n)$th
and the $(2+n)$th row to become the second and third, respectively, then
does likewise to the $(1+n)$th and $(2+n)$th column and thus generates the
$3\times 3$ block $T_3$ and so forth. As $n$ decreases towards 1 we meet
all the
\begin{equation}
  \widehat{P}_{N,n}^2(k) = T_{m+1}^{N-mn} \cdot T_m^{(m+1)n-N}
  \qquad\mbox{for} \quad \frac{N}{m}\geq n\geq\frac{N}{m+1}
  \label{F30a}
  \end{equation}
with $m=1,\ldots N$.

Simplest to deal with is, of course, the case of the smallest
non-trivial dimension, $N=2$, and there we obtain
\begin{equation}
  \widehat{P}_{2,1}^2(k) = J_0^2(k) + J_1^2(k),\qquad
  \widehat{P}_{2,2}^2(k) = J_0^2(k).
\label{F31}
\end{equation}
By Fourier transforming we produce the densities of the first and second
trace,
\begin{equation}
P_{2,1}^2(t)=\frac{\sqrt{4-|t|^2}}{2\pi^2|t|}, \qquad
P_{2,2}^2(t)=\frac{1}{\pi^2|t|\sqrt{4-|t|^2}} .
\end{equation}

Now we propose to show that the
distribution $P^2_{N,n}$ of the $n$th trace behaves like a Gaussian
with respect to its moments $\overline{|t_n|^{2m}}$ for sufficiently low
orders. To this end we define the functions
\begin{equation}
  \stackrel{\sim}{J}_m(k) = (-\i k)^m J_m(k)
\label{F32}
\end{equation}
which have the property (Abramowitz and Stegun 9.1.30 \cite{abr})
\begin{equation}
  \frac{\i}{k}\, \frac{\partial}{\partial k} \stackrel{\sim}{J}_m(k) = \,
  \stackrel{\sim}{J}_{m-1}(k)\,.
\label{F33}
\end{equation}
The Toeplitz determinant $T_N$ can then be rewritten as
\begin{equation}
  T_N = \det
  \left(\stackrel{\sim}{J}_m,\stackrel{\sim}{J}_{m-1},\ldots\right),\qquad
  \mbox{with}\quad m=0,\ldots,N-1.
  \label{F34}
\end{equation}
From this we find with (\ref{F32})
and (\ref{F33}) for $l\leq N$
\begin{equation}
  \left(\frac{1}{k}\,\frac{\partial}{\partial k}\right)^l
  T_N\left|_{k=0}\, =
  \left(\frac{-1}{2}\right)^l\right.\,.
\label{F35}
\end{equation}
To prove the foregoing identity we proceed as follows. Applying
$\frac{1}{k}\,\frac{\partial}{\partial k}$ once to $T_N$ we get a
determinant differing from $T_N$ only in the last column where according
to (\ref{F33}) $\stackrel{\sim}{J}_{m-N+1} \rightarrow
\stackrel{\sim}{J}_{m-N}$. Now setting $k=0$ and invoking
$\stackrel{\sim}{J}_m(0)=0$ for positive integer $n$ while
$\stackrel{\sim}{J}_m(0)=(\frac{-\i}{2})^m/(-m)!$ for $m=0,-1,-2,\ldots$
we face a determinant with vanishing elements below the diagonal and
thus equalling the product of its diagonal elements; among these only
the last, $\stackrel{\sim}{J}_1(0)=-\i /2$, differs from unity whereupon
(\ref{F35}) is proven for $\ell=1$. In the next step, $\ell=2$, $T_N$ is
changed such that only the elements in the last 2 columns  may get
their indices
shifted; again setting $k=0$ we are left with a $2\times 2$ determinant
which yields (\ref{F35}) for $\ell=2$. In the $\ell$th step we get a sum
of $\ell\times\ell$ determinants which is evidently independent of the
dimension $N$ for $\ell\leq N$. Due to that independence we need not
pursue the non-trivial task of proving (\ref{F35}) for arbitrary
$\ell\leq N < \infty$ but rather invoke the much more easily proven
result, to be established in the next section, that for
$N\rightarrow \infty$
the determinant $T_N$ is a Gaussian in $k$. It follows that for finite
$N$ the expansion of $T_N$ in powers of $k^2$ coincides
with that of
$\exp\left(-\frac{k^2}{4}\right)$ up to the $N$th power. Equations 
(\ref{F27}) - (\ref{F30}) show that $P_{N,n}^2(t)$ behaves
like a Gaussian distribution with respect to
all moments $\overline{|t_n|^{2m}}$ of orders $m\leq N/n$.
In particular, we find for the variances of the traces
\begin{equation}
  \overline{|t_n|^2} = \left\{
  \begin{array}{rll}          & n &  {\rm for}\quad 0 < n \leq N\\
                              & N &  {\rm for}\quad n \geq N .
  \end{array}
  \right.
\label{F37}
\end{equation}
Interestingly, these variances grow towards the Poissonian value $N$ as
$n\rightarrow N$ from below and then remain stuck as $n$ grows further.

We now proceed to the orthogonal and symplectic
cases. Starting as in the previous section with integrating over
alternating variables we find for the
orthogonal case with even $N$ and the symplectic case
\begin{eqnarray}
  \widehat{P}_{N,n}^1(k)\,\,&\propto&\,(\det A)^{1/2}
  \qquad\mbox{for}\quad\beta=1\nonumber\\
  \widehat{P}_{N,n}^4(k)\,\,&\propto&\,(\det B)^{1/2}
  \qquad\mbox{for}\quad\beta=4
\label{F39}
\end{eqnarray}
with the antisymmetric matrices
\begin{eqnarray}
\fl  A_{mm'}&=& \sum_{ss'} J_{|s|}(k) J_{|s'|}(k)\,
     \frac{(-\i)^{|s|}(-\i)^{|s'|}}{m+ns}\,
     \delta\left(m+m'+n(s+s')\right)
    \quad\mbox{with } |m|,|m'| \leq \frac{N-1}{2} \nonumber\\
\fl B_{mm'} &=& \sum_{s} J_{|s|}(2k)\,(-\i)^{|s|}\,
  \delta(m+m'+ns)
  \quad\mbox{with } |m|,|m'| \leq \frac{2N-1}{2} .
\end{eqnarray}
We have not proven the Gaussian property
but have calculated the variances. In the orthogonal case $(\beta = 1)$ we find
\begin{equation}
  \overline{|t_n|^2} = \left\{\begin{array}{rllll}
  &2n&-n\sum_{m=1}^{n} \frac{1}{m+(N-1)/2}\quad &\rm{for}& 0 < n \leq N
\label{F42} \\
   &2N&-n\sum_{m=1}^N \frac{1}{m+n-(N+1)/2}\quad
   &\rm{for}& \quad \quad n \geq N\,,
   \end{array}
   \right.
\end{equation}
while the symplectic case $(\beta= 4)$ yields
\begin{equation}
  \overline{|t_n|^2}  = \left\{\begin{array}{rllll} &\frac{n}{2}& +
  \frac{n}{4} \sum_{m=1}^{n}
  \frac{1}{N + \frac{1}{2}-m} \qquad &{\rm for}\quad 0<n\leq 2N \\
\label{F43}
   &N& \qquad \hspace{2.9cm}&{\rm for}\hspace{1.2cm} n\geq 2N\,.
  \end{array}
  \right.
\end{equation}

Needless to say, these mean squared traces could have been read off the
well-known two-level correlation functions of the circular ensembles the
Fourier transforms of which our variances in essence are \cite{mehta}.
Indeed, by introducing a non-normalized density of eigenphases as
\begin{equation}
  \rho(\varphi)=2\pi\sum_{i=1}^N \delta(\varphi-\varphi_i)=
                    \sum_{i=1}^N
                    \sum_{n=-\infty}^{\infty}\e^{-\i n(\varphi-\varphi_i)}
\end{equation}
one immediately sees that the two-point correlation function of that
density reads
\begin{equation}
   \overline{\rho(\varphi)\,\rho(\varphi')}= \sum_{n=-\infty}^{\infty}
   \overline{|t_n|^2}\,\e^{-\i n(\varphi-\varphi')}\,.
\end{equation}

\begin{figure}
\unitlength1cm
\begin{picture}(8,5)
\epsfxsize8cm
\put(1,\putoffset){\rotate[r]{\epsfbox{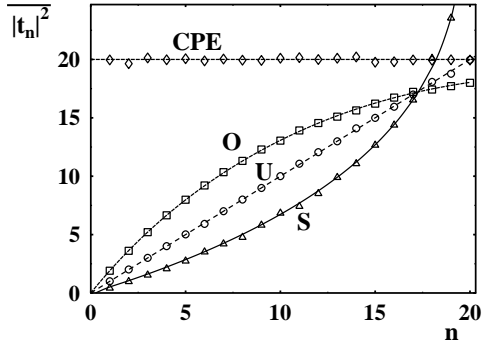}}}
\end{picture}
\unitlength1bp
\caption{Mean squared traces $\overline{ |t_n|^2}$ for $10^4$
  matrices of size $N=20$ typical of CPE $(\Diamond)$, COE (\opensqr),
  CUE (\protect\opencirc) and CSE
  $(\triangle)$ compared with analytical results (lines).}
\label{f4}
\end{figure}
Figure \ref{f4} reveals excellent agreement of the mean squared traces
as computed for samples of $4\times 10^4$ $20 \times 20$
matrices
of the four ensembles considered with the analytical predictions.
Note that for the
Poisson circular ensemble $\overline{|t_n|^2}$ equals the
matrix size $N$, without dependence on $n$. For small $n$ the variance
of traces decreases with the repulsion parameter $\beta$. The
data for the symplectic ensemble are obtained with $2N$
dimensional matrices, which
provide $N$ different eigenvalues each.

\begin{figure}
\unitlength1cm
\begin{picture}(8,5)
\epsfxsize8cm
\put(-1,\putoffset){\rotate[r]{\epsfbox{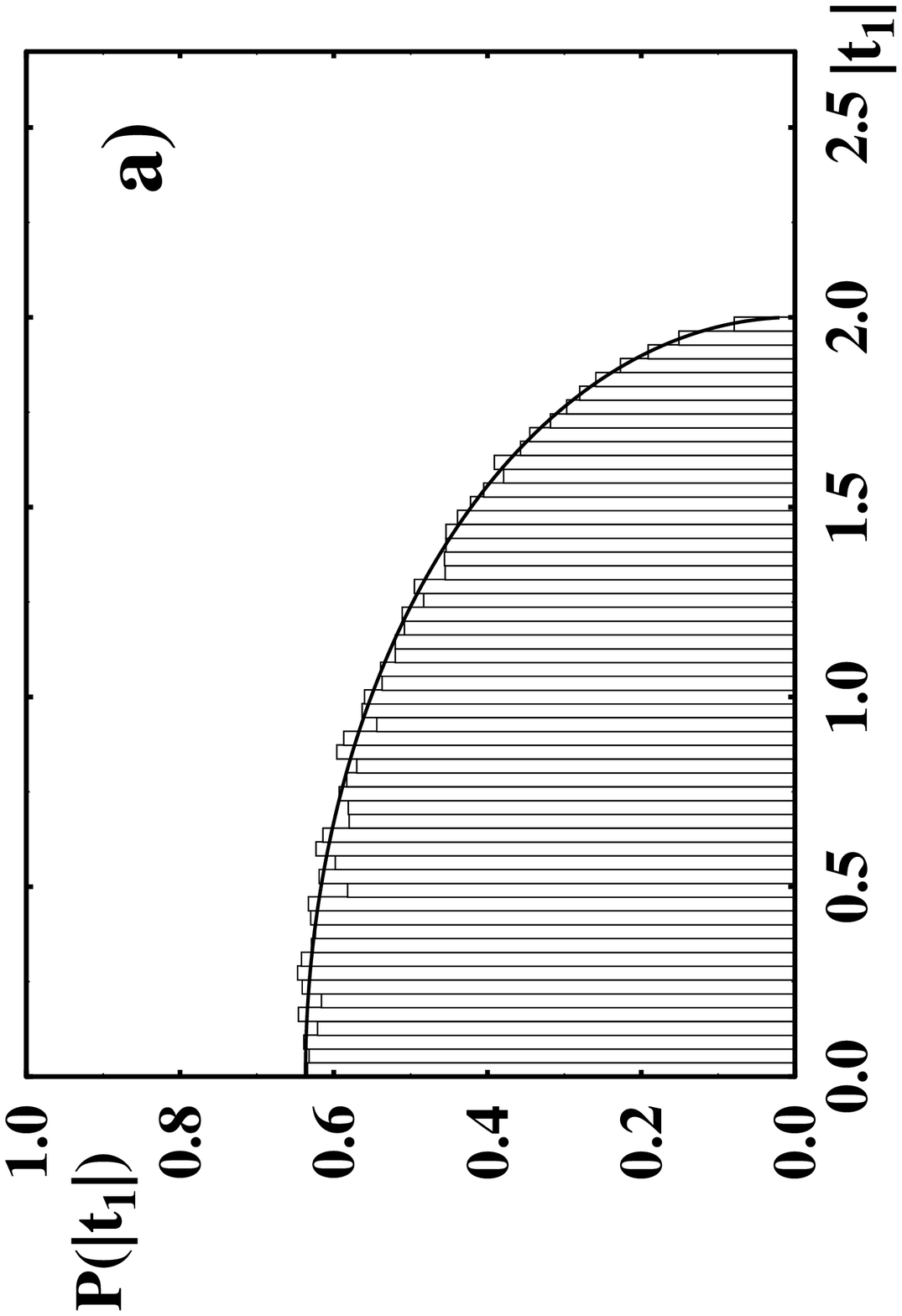}}}
\put(6.5,\putoffset){\rotate[r]{\epsfbox{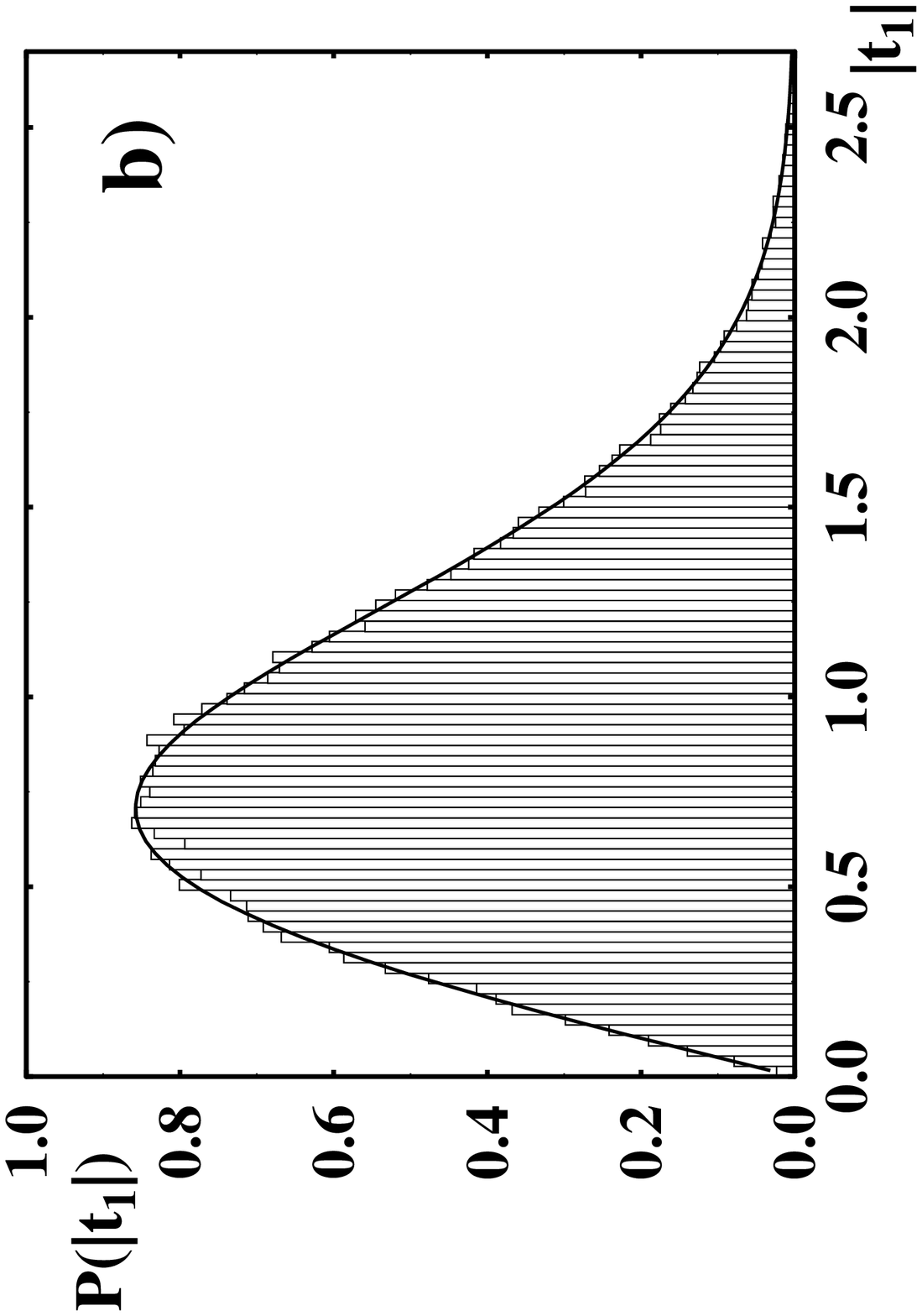}}}
\end{picture}
\unitlength1bp
\caption{Distribution of the modulus of the first trace $P(|t_1|)$ for
  $10^5$ CUE matrices for a) $N=2$ (semicircle distribution), and b)
  $N=20$ (Gaussian distribution).}
\label{f5}
\end{figure}
\begin{figure}
\unitlength1cm
\begin{picture}(8,5)
\epsfxsize8cm
\put(1,\putoffset){\rotate[r]{\epsfbox{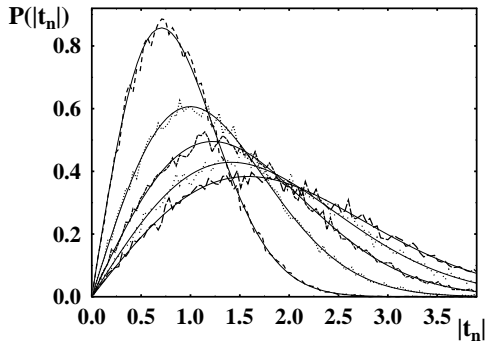}}}
\end{picture}
\unitlength1bp
\caption{Distribution of the moduli of the first five traces $|t_1|,
  \dots,|t_5|$ for $10^5$ CUE matrices of size $N=20$. Narrower solid
  lines correspond to Gaussian distributions with appropriate variances
  $\overline{|t_n|^2}=n$.}
\label{f6}
\end{figure}
Figures \ref{f5}, \ref{f6} display the similarly fine agreement of our
analytical results for the densities $P^{2}_{N,n}$ with numerical
data for sets of random matrices drawn from the circular unitary
ensemble. We have performed similar numerical studies for the orthogonal
and symplectic ensembles, again finding Gaussian marginal distributions for
the traces of sufficiently large matrices.
Moreover, in all cases studied, the distribution
of the traces was isotropic, i.e.\ without any phase dependence.

We should add a word of intuitive explanation to the statistics
of the traces for large dimensions $N$. The $n$th trace of a unitary
matrix may be thought of as a random walk in the complex plane, with
each of the $N$ steps of unit length and the $i$th step in a direction
given as $n$ times the $i$th eigenphase $\varphi_i$. These directions are
mutually independent and uniformly distributed for the Poissonian
ensemble which fact explains the independence of the characteristic
function (\ref{F24}) of $n$ and the ensuing first and second moments,
$\overline{t_n}=0,\quad\overline{|t_n|^2}=N$. In accordance with the
central limit theorem the rescaled traces $t_N/\sqrt{N}$ tend,
with $N\rightarrow\infty$, to have a
Gaussian distribution of zero mean and unit variance. Without such
rescaling the moments $\overline{|t_n|^{2m}}$ with $m\ll N$ differ
from those of the Gaussian defined by the first two moments only by
corrections of relative order $1/N$.

For the circular ensembles with $\beta>0$ the phases $\varphi_i$ display
repulsion of degree $\beta$ such that the directions of subsequent steps
in the random walk mentioned are not independent. The correlations
between the phases cannot prevent near-Gaussian behaviour of the traces
$t_n$ with $n\ll N$, as is intuitive in view of the local character of
the spectral correlations. Moreover, while the phases $\varphi_i$ cover the
interval $[0,2\pi)$ uniformly once, their multiples $n\varphi_i$ go around
that interval $n$ times such that for $n>1$ the phases
$n\varphi_i[\mbox{mod}\,(2\pi)]$
may exhibit accidental close neighborhoods of originally distant
$\varphi_i$.
\section{Joint density of traces for large CUE matrices}
We shall here employ a powerful theorem about determinants of Toeplitz
matrices, due originally to Szeg\"o and Kac and extended by
Hartwig and Fisher \cite{HF}, to find the marginal and joint
distributions of the traces $t_n$ of CUE matrices in the limit of large
dimension $N$. As our starting point we recall the identity (\ref{F5})
for the CUE average of a symmetric function of all $N$ phases.
Assuming, moreover, that symmetric function to have the form of a
product we can pull the integral over the $m$th phase $\varphi_m$ into
the $m$th row of the determinant in (\ref{F5}) and thus express the
average as a Toeplitz determinant,
\begin{equation}
\fl \overline{\prod_{m=1}^N f(\varphi_m)} =
\det(f_m,f_{m-1},\ldots,f_{-N+1}) =T\,(\{f\})\,,\qquad m=0,1,\ldots,N-1 ,
\label{F44}
\end{equation}
the elements of which are given by the Fourier transform
\begin{equation}
  f_m=\int^{2\pi}_0 \frac{\d\varphi}{2\pi}\e^{-\i m\varphi}f(\varphi)
\label{F45}
\end{equation}
of the function $f(\varphi)$. We had incurred two examples in
(\ref{F5}),\,(\ref{F6}) and (\ref{F25}),\,(\ref{F26}).
The theorem in question says that for large $N$ the above determinant
is given by
\begin{equation}
  \mbox{ln}\,T\,(\{f\}) = Nl_0+\sum_{n=0}^{\infty}nl_nl_{-n}
\label{F46}
\end{equation}
where the $l_n$ are the Fourier coefficients of
$\mbox{ln}\,f(\varphi)$,
i.e.\ $\mbox{ln}\,f(\varphi)=\sum_{n=-\infty}^{\infty}\,
l_n\e^{\i n\varphi}$. The conditions the function $f(\varphi)$ must meet
for the above limiting form to hold are (i) $f(\varphi)\neq 0$ for
$0\leq\varphi<2\pi$, (ii) $\mbox{arg}f(2\pi)=\mbox{arg}f(0)$, (iii)
$\sum_{n=-\infty}^{\infty}|f_n|<\infty$, and
(iv) $\sum_{n=-\infty}^{\infty}|n||f_n|^2<\infty$; they are fulfilled in
all examples of interest here.

In a first application we return to the Toeplitz determinant
(\ref{F26}) which gives the density $P_{N,n}^2$ of the $n$th trace. The
only non-vanishing Fourier coefficients of
$\mbox{ln}f(\varphi)=-\i k\cos{n\varphi}$ are $l_{\pm n}= -\i k/2$
whereupon we get the Gaussian anticipated in the previous section,
\begin{equation}
  \widehat{P}_{N,n}^2(k) = e^{-nk^2/4}\quad \Longleftrightarrow\quad
  P_{N,n}^2(t)=\frac{1}{n\pi}e^{-|t|^2/n},
\qquad\mbox{for}\quad N\gg 1.
\label{F47}
\end{equation}

No more difficult to obtain is the joint density of the first $n$ traces
\begin{equation}
P_{N}^{\beta}(t_1,\ldots,t_n) =
\overline{\prod_{m=1}^n\delta^2\left(t_m-\sum_{i=1}^N
\e^{\i m\varphi_i}\right)}
\label{F48}
\end{equation}
since its Fourier transform $\widehat{P}_N^2(k_1,\ldots,k_n)$
is once
more of the form (\ref{F44}) with
\begin{equation}
  f(\varphi)=\exp{\left(-\frac{\i}{2}\sum_{m=1}^n\left(k_m
  \e^{-\i m\varphi}+k_m^*
\e^{\i m\varphi}\right)\right)}.
\label{F49}
\end{equation}
The non-vanishing Fourier coefficients of the logarithm of that
latter function are $l_{m}=-\i k_m/2,\,l_{-m}=-\i k_m^*/2$ with
$m=1,\ldots,n$. The theorem (\ref{F46}) thus yields, for $N\gg n$,
\begin{eqnarray}
\fl\widehat{P}_N^2(k_1,\ldots,k_n)=
\exp{\left(-\sum_{m=1}^n m\,|k_m|^2/4\right)}\nonumber \\
\lo\Longleftrightarrow
  P_{N}^2(t_1,\dots,t_n)=\frac{1}{n!\pi^n}
  \exp{\left(-\sum_{m=1}^n|t_m|^2/m\right)} ,
\label{F50}
\end{eqnarray}
i.~e.~the product of the marginal distributions of the first
$n$ traces. The result generalizes in an obvious way to the joint
density of an arbitrary set of finite-order traces. We thus conclude
that in the limit $N\rightarrow\infty$ the finite-order traces are
statistically independent and all have Gaussian distributions.

We can now briefly comment on the conditions of applicability of the
Hartwig-Fisher theorem given above. The first two of them are clearly
fulfilled here since $\i\ln f(\varphi)$ as given by (\ref{F49}) is real,
continuous and periodic. The third and fourth condition are met since
the
derivative $f^\prime(\varphi)=\sum_{m=-\infty}^\infty
\i mf_m\e^{\i m\varphi}$ is square integrable (trivially indeed since 
$\i \ln f(\varphi)$ is a finite Fourier series); in particular,
$f^\prime(\varphi)$
obeys Parseval's identity,
\begin{equation}
\sum_{m=-\infty}^\infty|m|^2|f_m|^2
=\int\limits_0^{2\pi}|f^\prime(\varphi)|^2\d\varphi=
\pi\sum_{m=1}^nm^2|k_m|^2 <\infty\;.
\label{F50a}
\end{equation}
Since  $|m||f_m|^2\le|m|^2|f_m|^2$ it follows that
$\sum_{m=-\infty}^\infty|m||f_m|^2<\infty$, i.\,e. the validity of
condition (iii).

Finally, we invoke Cauchy's inequality in
\begin{equation}
\fl \sum_{m=-\infty}^\infty|f_m|=|f_0|+ \sum_{m\ne
0}^\infty\left|\frac{1}{m}\right||mf_m| \leq |f_0|+\left(\sum_{m\ne
0}^\infty\left|\frac{1}{m}\right|^2\right)^{1/2}\left(\sum_{m\ne
0}^\infty|mf_m|^2\right)^{1/2}\;.
\end{equation}
Upon using (\ref{F50a}) and the convergence of $\sum_{m\ne
0}^\infty\frac{1}{|m|^2}$ we verify condition (iv).

For finite dimension the independence as well as the Gaussian character
of the traces are only approximate. For sets of traces both of these
properties tend to get lost as the sum of the orders of the traces in a
set increases. In particular, since all traces $t_n$ are uniquely
determined by the N real eigenphases, only $N/2$ traces can be
independent.

Preliminary numerical studies suggest that the finite-order traces might
be similarly independent and Gaussian for the COE and the CSE. For the
CPE, of course,
the independence holds trivially.

\section{Remarks on the distributions of coefficients of secular
polynomials }
\label{s:distann}

As we have just seen the first few traces $t_1,\,t_2,\ldots$ of large
unitary matrices drawn from any of the circular ensembles display no
noticeable correlations. It follows that the coefficients
$a_1,\,a_2,\ldots$ must bear strong mutual correlations simply
since $a_n$ can be expressed in terms of the first $n$ traces through
Newton's formulae (\ref{F19}). One must therefore expect that the explicit
form of the marginal and joint distributions of the $a_n$ are hard
to come by. An exception is provided by $a_1$ the marginal
distribution of which is trivially related to that of the first
trace since $a_1=t_1$.

A slightly less trivial result may be
obtained for the distribution $P(a_2)$ of the second coefficient:
Since $a_2=\frac{1}{2}(t_1^2-t_2)$ we  may invoke the CUE joint
distribution (\ref{F50}) of the first two traces to get
\begin{equation}
 P(a_2)= \sqrt{\frac{2e}{\pi^3}}\int_0^2\frac{\d x}{\sqrt{x(2-x)}}
 \exp{\left(-\frac{1}{x}-|a_2|^2x\right)} \,.
 \label{faltung}
 \end{equation}
A saddle-point approximation to the foregoing integral
immediately reveals that $P(a_2)$ decays exponentially for large $a_2$.
Proceeding similarly one may combine Newton's formulae with the joint
distribution of the traces to get the marginal and joint distributions
of the first few $a_n$, with decreasingly compact and enjoyable results.

\begin{figure}
\unitlength1cm
\begin{picture}(8,5)
\epsfxsize8cm
\put(1,\putoffset){\rotate[r]{\epsfbox{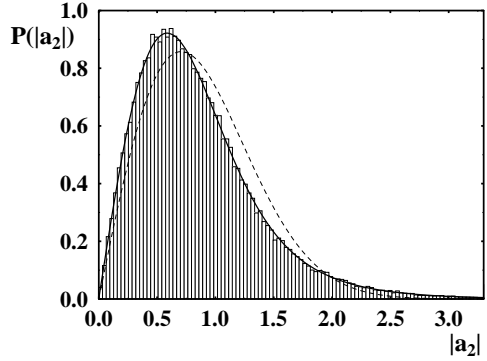}}}
\end{picture}
\unitlength1bp
\caption{Distribution of the moduli of second coefficients $|a_2|$
  for $10^5$ CUE matrices of size $N=20$. Solid line represents
  $2\pi|a_2|P(a_2)$ according to (5.1)
  and is compared to a dashed line representing Gaussian
  distribution of $|a_1|$.}
\label{f7}
\end{figure}
\begin{figure}
\unitlength1cm
\begin{picture}(8,5)
\epsfxsize8cm
\put(1,\putoffset){\rotate[r]{\epsfbox{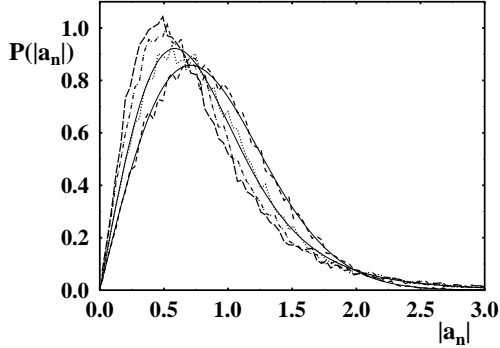}}}
\end{picture}
\unitlength1bp
\caption{Distribution of the moduli of the coefficients  $|a_1|$
  (short dashed line), $|a_2|$ (dotted), $|a_4|$ (dash-dotted)  and
  $ |a_6|$ (long dashed line) for CUE matrices of size $N=20$.
  Distributions of the two first coefficients are compared with
  theoretical predictions (narrow lines).}
\label{f8}
\end{figure}
Figure \ref{f7} presents the distribution of the modulus of
$|a_2|$ as obtained numerically from $10^5$ CUE matrices of the size $10
\times 10$. These numerical data agree well with the distribution
$2\pi|a_2| P(a_2)$ according to (\ref{faltung}). In Figure \ref{f8} we
depict numerical results for the distributions of the moduli of some higher
coefficients $|a_n|$. All of these curves grow linearly out of the
origin with a slope increasing with the index $n$ of the coefficient
$a_n$. On the other hand, all of these distributions but the first are
characterized by a long exponential tail. The latter originates from the
convolution type integrals which combine the densities of the traces
to those of the coefficients; it contrasts with the Gaussian tail of the
distribution of the first coefficient. The qualitative features just
outlined for the CUE are shared by the distributions $P(a_n)$ for
the other canonical ensembles.

\section{Comparison with a dynamical system}
\label{s:top}

We here propose to examine in how far our results on distributions of the
coefficients $a_n$ and traces $t_n$ of matrices drawn from the canonical
circular ensembles are applicable to dynamical systems. Choosing the
familiar model of the periodically kicked top \cite{haake}, \cite{hks87}
we work with a Floquet operator of the form
\begin{eqnarray}
  \fl F= \exp\left(-\i\frac{k_x}{2j+1}J_x^2-\i p_xJ_x\right)\,
  \exp\left(-\i\frac{k_y}{2j+1}J_y^2-\i p_yJ_y\right)\,
  \nonumber\\ \times \exp\left(-\i\frac{k_z}{2j+1}J_z^2-\i p_zJ_z\right)\,.
\label{toptop}
\end{eqnarray}
This involves the components of an angular momentum operator
$J_x,J_y,J_z$ which satisfy standard commutation relations,
$[J_x,J_y]=\i J_z$ etc. The quantum number $j$ fixes $\vec{J}^2=j(j+1)$
and the size of the Hilbert space, $N=2j+1$.

For generic values of the parameters $k_x, k_y, k_z$ and $p_x, p_y, p_z$
the corresponding classical dynamics is chaotic
and there is no geometric nor antiunitary symmetry left \cite{hks87}.
All previously studied statistical properties of the quasienergy
spectrum and the eigenvectors were found to be remarkably
faithful to the predictions of the CUE \cite{haake}.
On the other hand, when the parameters $k_y$ and $p_y$ (or, instead,
$k_x$ and $p_x$) are set
to zero, an antiunitary symmetry under time reversal appears, and in
that case the spectral and eigenvector statistics were found as of the
COE type. We shall refer to the two variants of the model as to the
unitary and orthogonal top, respectively.

Before presenting our data for the traces $t_n$ and the secular
coefficients $a_n$ of various tops a word of caution is in order.
Previous statistical analyses of tops were made for spectrally
local quantities like low-order correlations of the level density,
the distribution of nearest-neighbor spacings,
or for components of eigenvectors; what distinguishes these
quantities is a certain self-averaging character: A single
Floquet matrix of large dimension $N$ provides a sufficient data basis
to extract reliable means or even distributions from. Now, the sequence of
the traces $t_n$ and that of the secular coefficients $a_n$ are not in
any way self-averaging since such a sequence with $1\leq n\leq N/2$
stands for a whole quasienergy spectrum and therefore changes in a
system specific manner when control parameters are varied. Consequently,
it would not make sense to compare such a sequence for an individual
Floquet matrix with the means calculated here for the various circular
ensembles. We must rather ask whether an ensemble of Floquet matrices of
the type (\ref{toptop}), defined by a whole set of values for the various
control parameters, is faithful to the prediction based on the circular
ensemble of random matrices of the same symmetry class. It is in this
sense that we have undertaken the comparison to follow.

\begin{figure}
\unitlength1cm
\begin{picture}(8,5)
\epsfxsize8cm
\put(1,\putoffset){\rotate[r]{\epsfbox{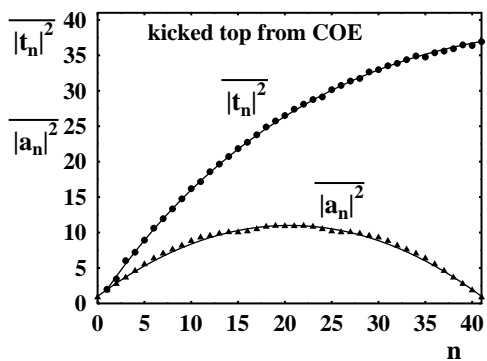}}}
\end{picture}
\unitlength1bp
\caption{Mean squared traces $\overline{ |t_n|^2}$ (circles)
  and coefficients $\overline{ |a_n|^2} $ (triangles) for
  orthogonal tops are compared with COE results (lines). Data
  are averaged over 20000 different Floquet matrices of the structure
  (6.1) as described in the text.}
\label{f9}
\end{figure}
Figure \ref{f9} shows the $n$-dependence of the variances of the traces
(circles) and of the coefficients (triangles) for orthogonal tops.
The order of the characteristic polynomial was taken as
$N=41$ by choosing $j=20$. Data from 20000 matrices were gathered by
picking $k_y, k_z$ from intervals of length 3 around 10 and $p_y, p_z$
from $[3\pi/8,5\pi/8]$, all with independent box distributions. These
intervals were chosen so as to secure classical chaos and to avoid
geometric symmetries. The agreement with the COE is obviously
satisfactory.

\begin{figure}
\unitlength1cm
\begin{picture}(8,5)
\epsfxsize8cm
\put(1,\putoffset){\rotate[r]{\epsfbox{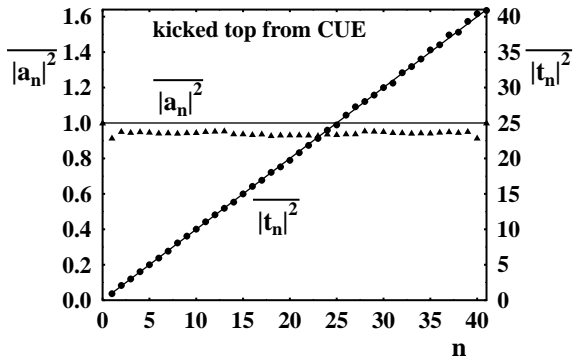}}}
\end{picture}
\unitlength1bp
\caption{As in Figure \protect\ref{f9} for unitary tops.
  Lines denote the CUE results.}
\label{f10}
\end{figure}
\begin{figure}
\unitlength1cm
\begin{picture}(8,5)
\epsfxsize8cm
\put(1,\putoffset){\rotate[r]{\epsfbox{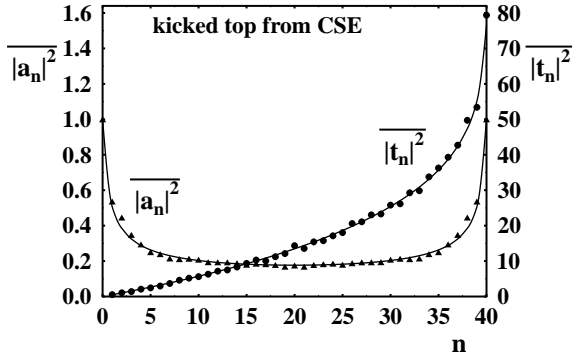}}}
\end{picture}
\unitlength1bp
\caption{As in Figure \protect\ref{f9} for symplectic tops. Average over
  2000 matrices as described in the text. Lines denote CSE results.}
\label{f11}
\end{figure}
Analogous data for unitary tops are presented in Figure \ref{f10}. The
agreement with the CUE predictions is acceptable in the sense that there
is no doubt about the universality class. However, system specific
behaviour is clearly visible for low-order traces and even in all secular
coefficients. Qualitatively, such deviations from random-matrix theory
are not unexpected since low-order traces do contain primarily system
specific information retrievable from short periodic orbits and since
the $a_n$ even for large $n$ contain low-order traces as expressed in
Newton's formulae.

Symplectic tops can also be constructed by securing an antiunitary
symmetry $T$ with $T^2=-1$ and avoiding geometric symmetries
\cite{haake}. The Floquet operator
\begin{eqnarray}
\fl
  F=\exp\left(-\i\frac{k_1}{2j+1}J_z^2-\i\frac{k_2}{2(2j+1)}(J_xJ_z+J_zJ_x)
              -\i\frac{k_3}{2(2j+1)}(J_xJ_y+J_yJ_x) \right)\,
  \nonumber\\  \times\exp\left(-\i\frac{k_4}{2j+1}J_z^2\right)
\end{eqnarray}
in a representation with halfinteger $j$ serves fine. A set of 2000
matrices with $j=39.5$ and coupling constants $k_i$ drawn at random from
a hypercube of length 0.4 near 10 gives data shown in Figure \ref{f11}.
The agreement with the CSE is better than could be hoped for.

%----------------------------------------------------------------------
\ifiop
\ack
\else
\section*{Acknowledgments}
\fi
We would like to thank Nils Lehmann for helpful cooperation and Pragya
Shukla for first efforts to calculate the mean squared secular
coefficients for the COE and the CSE. Financial support by the
Sonderforschungsbereich ``Unordnung und gro{\ss}e Fluktuationen'' der
Deutschen Forschungsgemeinschaft is gratefully acknowledged.

%----------------------------------------------------------------------
\appendix
\section{Differential equation for the generating function}

Let us recall the heuristic way we first went towards the mean squared
secular coefficients (\ref{F13}), (\ref{F17}), and (\ref{F18}). After
deriving the Pfaffians (\ref{F12a}) and (\ref{F15}) we applied an
algebraic computer program to evaluate the $\overline{|a_n|^2}$ for
small dimensions $N$ and then proceeded analytically to $n=1,2$ and
arbitrary $N$. We thank N Lehmann for the computer work and also for his
intuition in guessing (\ref{F17}) with us. Having eventually guessed
(\ref{F13}) as well we speculatively extrapolated to the general form
(\ref{F18}) for arbitrary positive $\beta$.

To prepare for the proof of (\ref{F18}) we observe the equivalence to
the recursion relation
\begin{equation}
  (n+1)\left(1+\frac{\beta}{2}(N-n-1)\right)
  \overline{|a_{n+1}|^2}=(N-n)\left(1+n\frac{\beta}{2}\right)
  \overline{|a_{n}|^2}
\label{A1}
\end{equation}
with $\overline{|a_{0}|^2}=0$. Multiplying (\ref{A1}) by $x^n$, using
$(x\frac{\partial}{\partial x})x^n=nx^n$, and summing from $n=0$ to
$n=N-1$ we obtain the differential equation (\ref{F19}) the polynomial
solution of which with $P_N^{\beta}(x)=1$ generates the mean squared
secular coefficients. A few elementary steps will now yield the validity
of the differential equation.

We start from the definition (\ref{F2}) of the generating function,
setting $\lambda=x,\, \mu=1$ and take the derivative with respect to $x$,
\begin{equation}
  \frac{\partial P(x)}{\partial x} = -\left\langle \sum_{i=1}^N
  \frac{1}{\e^{\i\varphi_i}-x} \right\rangle\, ,
\label{A2}
\end{equation}
where the angular brackets denote an average with the whole integrand of
(\ref{A2}) as the weight. On the other hand, taking the derivative with
respect to $\mu$ before setting $\mu=1$ and exploiting the invariance of
the $N$-fold phase integral under a constant shift of all phases we get
\begin{equation}
  x\frac{\partial P(x)}{\partial x} = -\left\langle \sum_{i=1}^N
  \frac{1}{\e^{-\i\varphi_i}-1} \right\rangle\, .
\label{A3}
\end{equation}
Next, we differentiate (\ref{A2}) once more,
\begin{eqnarray}
  \frac{\partial^2 P(x)}{\partial x^2} &=& \left\langle \sum_{i\neq j}
 \frac{1}{(\e^{\i\varphi_i}-x)(\e^{\i\varphi_j}-x)}\right\rangle\nonumber\\
  &=& \left\langle \sum_{i\neq j}
  \frac{-2}{(\e^{\i\varphi_i}-x)(\e^{\i\varphi_i}-\e^{\i\varphi_j})}
  \right\rangle\nonumber\\
  &=&-\left\langle \sum_{i\neq j}
  \frac{\e^{-\i\varphi_i}}{(\e^{\i\varphi_i}-x)}
  \left(1-\i\cot{\frac{\varphi_i-\varphi_j}{2}}\right)\right\rangle
  \,.
\label{A4}
\end{eqnarray}
Realizing that the cotangent function may be replaced by a derivative
acting on the joint density of eigenvalues (\ref{d}),
\begin{equation}
  \left(\frac{\partial}{\partial \varphi_i}-\frac{\beta}{2}
  \sum_{j(\neq i)} \cot{\frac{\varphi_i-\varphi_j}{2}} \right)d_N^{\beta}
  =0 \, ,
\label{A5}
\end{equation}
we can transform the last member in (\ref{A4}) by partial integration.
A little algebraic hocuspocus then gives, with the help of (\ref{A2})
and (\ref{A3}),
\begin{eqnarray}
  \fl \frac{\partial^2 P(x)}{\partial
  x^2}=\left(N-1+\frac{2(2-x)}{\beta(1-x)}\right)\frac{1}{x}\left(
  \frac{\partial P}{\partial
  x}+\sum_i\left\langle \e^{-\i\varphi_i}\right\rangle\right)
  \nonumber\\
  +\frac{2}{\beta(1-x)}
  \left(-\sum_i\left\langle
\e^{-\i\varphi_i}\right\rangle-NP+x\frac{\partial P}{\partial x}\right) \,.
\label{A6}
\end{eqnarray}
Finally, the average $\sum_i\langle \e^{-\i\varphi_i}\rangle$ can be
expressed in terms of the generating function $P$ and its derivative
$\partial P/\partial x$ by employing the identity $\sum_k\langle
\e^{-\i\varphi_k}\rangle=\sum_k\langle \i\frac{\partial}{\partial\varphi_k}
\e^{-\i\varphi_k}\rangle$, integrating by parts, and again invoking
(\ref{A2},\ref{A3},\ref{A5}),
\begin{equation}
  \sum_i\left\langle
  \e^{-\i\varphi_i}\right\rangle\left(2+\frac{\beta}{2}(N-1)\right)=(x-1)
  \frac{\partial P}{\partial x} -NP\,.
  \label{A7}
\end{equation}
Upon inserting this in (\ref{A6}) we obtain a differential equation for
the generating function $P$ which is easily put into the form (\ref{F19}).
%----------------------------------------------------------------------
\newpage
\ifiop
  
\else
  \end{references}
\fi

\section*{References}
  \begin{thebibliography}{99}
\else
  \begin{references}
\fi
\bibitem{dys62} Dyson F J 1962 {\sl J.\ Math.\ Phys.\ }{\bf 3} 140
\bibitem{mehta} Mehta M L 1991 {\sl Random Matrices} II ed.
  (New York: Academic)
\bibitem{haake} Haake F 1990 {\sl Quantum Signatures of Chaos}
  (Berlin: Springer)
\bibitem{jp95} Jalabert R A and Pichard J-L 1995 {\sl J.\ Phys.\ I France}
  {\bf 5} 287
\bibitem{ga70} Gaudin M 1966 {\sl Nuclear Physics} {\bf 85} 545
\bibitem{ps91} Pandey A and Shukla P 1991 {\sl J.\ Phys.\ }A {\bf 24} 3907
\bibitem{mi95} Muttalib K A and Ismail M E H 1995 {\sl J.\ Phys.\ }A
  {\bf 28} L541
\bibitem{lewd91} Lewenkopf C H and Weidenm\"uller H A 1991
  {\sl Ann.\ Phys.}, NY {\bf 212} 53
\bibitem{bro95} Brouwer P W 1995 {\sl Phys.\ Rev.\ }B {\bf 51} 16878
\bibitem{pm83}  Pereyra P and Mello P A 1983 {\sl J.\ Phys.\ }A
  {\bf 16} 237
\bibitem{mps85} Mello P A, Pereyra P and Seligman T H 1985 
  {\sl Ann.\ Phys.}, NY {\bf 161} 254
\bibitem{dsf91}  Doron E, Smilanski U and Frenkel A 1991
  {\sl Physica} D {\bf 50} 367
\bibitem{bm94}  Baranger H U and Mello P A 1994 {\sl Phys.\ Rev.\ 
  Lett.\ }{\bf  73} 142
%\bibitem{m90} Mello P A 1990 {\sl J.\ Phys.\ }A {\bf 23} 4061
\bibitem{Newton} Mostowski A and Stark M 1964 {\sl Introduction to Higher
  Algebra} (Oxford: Pergamon Press)
\bibitem{lbb92}  Bogomolny E, Bohigas O and Leboeuf P 1992
  {\sl Phys.\ Rev.\ Lett.\ }{\bf 68} 2726
\bibitem{keh93}
  Ku\'s M, Haake F and Eckhardt B 1993 {\sl Z.\ Phys.\ }B {\bf 92} 221
\bibitem{lbb95}  Bogomolny E, Bohigas O and Leboeuf P 1995 preprint
\bibitem{kzmk94} \.Zyczkowski K and Ku{\'s} M 1994 {\sl J.\ Phys.\ }A
  {\bf 27} 4235
\bibitem{kz95} \.Zyczkowski K 1995 {\sl Chaos - The Interplay between
  Stochastic and Deterministic Behavior (Proc.\ of
  XXXIst Winter School of Theoretical Physics: Karpacz, February 1995)}
  ed P Garbaczewski, M Wolf and A Weron (Springer) p 565
\bibitem{kzmk95} \.Zyczkowski K and Ku{\'s} M 1996 {\sl Phys.\ Rev.\ }E
  {\bf 53} 319 
\bibitem{abr} Abramowitz M and Stegun I A (eds.) 1970
  {\sl Handbook of Mathematical Functions} (New York: Dover Publ.)
\bibitem{HF} Hartwig R E and Fisher M E 1969
  {\sl Arch.\ Rat.\ Mech.\ Anal.\ }{\bf 32} 190
\bibitem{hks87} Haake F, Ku{\'s} M and Scharf R 1987
  {\sl Z.\ Phys.\ }B {\bf 65} 381
\ifiop
  \end{thebibliography}
\end{document}